# Enhancement of the response of non-uniform resonance modes of a nanostructure in the Picoprobe microwave-current injection ferromagnetic resonance

C.S.Chang[1], A.O.Adeyeye[2], M.Kostylev[1] and S. Samarin[1]

[1]*School of Physics, University of Western Australia, Australia;*
[2]*Department of Electrical and Computer Engineering, National University of Singapore, Singapore*

Abstract: The non-uniform standing spin-wave modes in thin magnetic films and nanostructures provide important information about surfaces and buried interfaces. Very often they are lacking in the recorded ferromagnetic resonance spectra for symmetry reasons. In this work we experimentally demonstrate that by direct injection of microwave currents into an array of Permalloy nanostripes using a microscopic microwave coaxial to coplanar adaptor one can efficiently excite non-uniform standing spin wave modes with odd symmetry. The proposed method is quick and allows easy spatial mapping of magnetic properties with the resolution down to 100 microns. We have validated this method using an example from a periodical array of nanostripes. The results from direct current injection are compared to that of microstrip-based FMR measurements.

Standing spin wave modes (SSWMs) are microwave magnetic excitations in confined geometries. The wavelengths of SSWMs are determined by the size of the sample along the confinement direction and pinning at the surfaces and interfaces. It is well known that the homogeneous microwave magnetic field typically used for ferromagnetic resonance (FMR) cavity experiments does not allow SSWM observation unless pinning [1] of magnetization is present at the sample surfaces.

The standing wave modes are affected by the inhomogeneous exchange interaction, and thus carry important information about surfaces and buried interfaces [1-4], and about the value of exchange constant for the material, so the possibility of observation of these modes using such a simple tool as FMR is very exciting. Recently, it was shown that in the microstrip based broadband FMR experiment [5-7], it is possible to considerably increase the FMR response of the higher-order SSWM modes in conducting ferromagnetic films, due to injection of eddy currents into the samples. However, in the microstrip FMR the efficiency of standing-wave mode excitation is strongly dependent on the driving frequency; in Permalloy (Py) films at the frequencies below 6 GHz, the higher-order standing wave modes show very small responses, if any. Furthermore, macroscopic-size coplanar or microstrip transducers (about 1mm in width, several mm long microstrip lines) are typically required in order to excite and observe SSWM. A way to get around this deficiency has recently been shown by Khivintsev et al [18], where the magnetic sample is embedded into a microscopic microwave coplanar transmission line. It has been demonstrated that if a magnetic sample is in the form of a stripe of a *microscopic* cross-section is sandwiched between two highly conducting (copper) layers and the sandwich forms the signal line of the coplanar waveguide, efficient absorption by the first anti-symmetric standing spin wave mode is seen in the transmission (S21) characteristic of the coplanar line. The microwave currents flowing in the Copper layers create an Oersted field which is anti-symmetric across the thickness of the magnetic layer. This forms the necessary conditions for excitation of this mode.

In this letter, we demonstrate that efficient excitation of an exchange standing spin-wave mode can be achieved in a much simpler way, without embedding the sample in a characterization gear. This gives more flexibility and can be used for determining the quality of nanomaterials. Furthermore, in contrast to Ref.[8], we demonstrate this effect in patterned *nano*structured material. The method is based on injection of microwave currents *directly* into a Permalloy nanostructure using the commercially available Picoprobe®; a microscopic microwave coaxial to coplanar adaptor. The method is quick and allows easy spatial mapping of magnetic dynamics with the resolution down to 100 microns, which is given by the minimum lateral size of the tip of the picoprobe. Both responses of the fundamental (quasi-uniform) dipolar modes and of non-uniform exchange modes are seen in this arrangement.

The objects of our study are arrays of magnetic nanostripes (MNS). These nanostructures are promising for magnonic [9] and magneto-plasmonic applications [10]. In the latter case, covering magnetic material with a thin gold (Au) layer, thus forming an Au/Py interface, may enhance structure performance [11]. A number of periodic arrays of parallel Permalloy (Py) nanostripes which are 100 nm thick, 264 nm wide, 4 mm long and with different inter-stripe spacings have been fabricated using deep ultra-violet lithography [12]. Single-layer Py samples and samples having a 10nm-thick Au capping layer have been prepared on Silicon substrates. A number of reference continuous films with the same thickness and composition have also been fabricated in the same process. All MNS and films demonstrate similar behavior, therefore in the following we concentrate on one single-layer MNS sample with an inter-stripe spacing of 150 nm and the respective reference film.

The tip of a picoprobe represents a set of three needle-like tungsten microwave contacts aligned in a row (see Fig. 1(a)). The central contact is connected to the central conductor of the feeding coaxial line and the outer ones are grounded. The distance from the central contact to the external ones is called "pitch". Below we demonstrate results obtained with a picoprobe having a pitch of 200micron. We saw similar behavior with a picoprobe having the smallest commercially available pitch: 50micron.

To drive magnetization precession and to register FMR absorption, contacts of the picoprobe are placed on top of the nanostructure or the film such that a microwave current from the contacts is directly injected into Py. In the case of MNS the sample is oriented in such a way that the line in the *z*-direction connecting the tips of all three contacts of the picoprobe ("tip line") is parallel to the applied static field and to the nanostripes. Thus, a microwave conduction current can flow between the picoprobe contacts. In the case of the continuous film, care is only taken that the applied field is oriented along the tip line such that the field is parallel to the microwave current and thus perpendicular to the microwave magnetic field of the current.

The tip of the picoprobe is lowered using a 3D translation stage. We monitor the approach of the picoprobe's tip to the film surface with a digital microscope. Electrical contact is monitored using a dc Ohmmeter. In full electrical contact of the picoprobe with the material, the measured dc picoprobe resistance is typically around 6 Ω and 130 Ω for the contacts with the continuous film and MNS respectively.

We measure the microwave signal reflected from the picoprobe as a function of the applied magnetic field $H_a$ for given microwave frequencies. Similar to [13], we modulate the magnetic field applied to the sample using two small coils attached to the poles of an electromagnet. Modulation frequency is 220Hz and the RMS magnetic field produced by the coils is 9.5 Oe. The microwave power in the frequency range 2-18 GHz is applied to the picoprobe from a microwave generator through a circulator. Signal reflected from the picoprobe is rectified using a microwave detector and detected using a digital lock-in amplifier referenced by the same 220 Hz signal. The signal we observe this way represents the first derivative of the resonance line (Fig. 2).

In a similar manner, microwave absorption in a microstrip FMR is detected. We use a section of a microstrip line 0.3 mm wide and 5 mm long (Fig.1 (b)). The sample is placed on top of the microstrip. MNS are oriented along the microstrip and along the applied field (axis *z*). FMR absorption is detected in the same way as in the picoprobe experiment, with the exception that a circulator was not used; the measured signal is that which is transmitted through the microstrip line.

Fig. 2(a) displays the results we obtained on the nanostripe sample. One observes two well-resolved modes, and one partially-resolved mode which is visible as an additional small peak on the lower-field shoulder of the lower-field well-resolved mode. The amplitudes of well-resolved higher and lower field modes swap between the picoprobe and the microstrip experiments which is the main finding in this work.

In order to explain this effect, one first has to identify these modes. This is easily done using the theory from [14,15]. It states that the resonance frequencies of the structured material should obey the approximate dispersion relation for spin waves valid for continuous films. All peculiarities of confinement due to nanostructuring is hidden in the coefficients of this equation. We write down this equation in the form

$$(\omega/\gamma)^2 = (H_a + H_{ex} + 4\pi M - H_d)(H_a + H_{ex} + H_d), \quad (1)$$

where $H_a$ is the applied field, $H_{ex}$ is the effective exchange field, $H_d$ is the effective dynamic dipole field, $M$ is the saturation magnetization of the material, ω is the spin wave angular frequency, and $\gamma/2\pi$ =2.82 MHz/Oe is the gyromagnetic coefficient. We fitted the experimental data for ω vs $H_a$ (Fig. 3) and extracted the values of $M$, $H_{ex}$, and $H_d$ for both modes. For the higher field mode, we obtained $H_{ex}$ =369 Oe and $H_d$ = 0 Oe while for the lower field mode we get $H_{ex}$ = 409 Oe and $H_d$ = 1418 Oe.

Based on the large value of $H_d$, the lower field mode is identified as the fundamental (dipole) mode of the array. This identification is confirmed by a numerical simulation using theory from [16]. The mode's resonant field is strongly shifted down field due to confinement effects resulting in strong effective magnetization pinning at the stripe edges [15] and a large dynamic demagnetizing (dipole) field. Similarly, from the vanishing $H_d$, the higher field mode is identified as the first exchange mode for MNS. This identification is also confirmed by the numerical simulation. The simulation shows that this mode has a simple (quasi-uniform) distribution of dynamic magnetization in the array plane but an anti-symmetric distribution across the stripe thickness. It represents the counterpart of the 1[st] exchange SSWM for the continuous film. The dipole field is small for this mode because of its anti-symmetric character [17]. Thus the main contribution to the mode frequency originates from the exchange energy which depends mainly on the smallest dimension of the structure; in our case, on the thickness of the nanostripes. Since the MNS thickness is the same as the thickness of the reference continuous film, one may expect that the resonant field for this mode should be close to the resonant field for the 1[st] SSWM of the film.

Fig. 2 (b) demonstrates the results obtained on the reference continuous sample at 14 GHz. One sees that the microstrip data contain one main response which is easily identified as the fundamental (uniform) FMR mode, and a small – barely observable – mode down-field from the fundamental mode, which has the frequency close to the frequency of the upper-field mode for MNS. The small-amplitude mode is identified as the first (anti-symmetric) SSWM. The extremely small higher-order response, also when the film is flipped such that film substrate faces the microstrip transducer [7] (not shown), demonstrates that the film has nearly perfect unpinned surface spins. The same fact is also confirmed by measurement with a cavity at 9.45 GHz. The gradual enhancement of the absorption amplitude for first SSWM mode's at larger frequencies is consistent with the increase in the effect of the eddy current injection at higher frequencies in the microstrip FMR [5].

This measurement taken on the reference film confirms that the higher field resonance peak for MNS is the 1$^{st}$ exchange mode. Thus, the effect we see in Fig. 2(a) is enhancement of the signal of the first SSWM in the picoprobe experiment with respect to the microstrip one. Interestingly, we do not observe a noticeable effect of this type for the continuous film. The picoprobe data (Fig.2b) show a small peak at the lower field shoulder of the fundamental mode consistent with the first SSWM. The ratio of the amplitude of the first SSWM mode to the fundamental mode is not significantly increased. Furthermore, the overall resonance amplitude obtained using the picoprobe on the continuous film is about twenty times smaller compared to the stripline method (Fig. 2b).

We see that in the confined MNS geometry the direct injection of the microwave currents enhances excitation of the anti-symmetric thickness modes with respect to the microstrip FMR, but no enhancement is seen for the continuous film. Below we suggest an explanation for this behavior.

In the microstrip based FMR, magnetization precession is excited by the in-plane component of the microwave magnetic field which is along $x$ in Fig. 1. This fact is evidenced by rotating the microstrip by 90 degree with respect to the applied field, resulting in a significantly diminished FMR signal. In the picoprobe-based experiment, the microwave voltage applied between the tips of the picoprobe induces a microwave current between the tips (along $z$). The Oersted field (along $y$) of this current drives magnetization precession. If the current density were uniform across the sample thickness (along $y$), the Oersted field would be an anti-symmetric function of $y$ with a zero at half a sample thickness. Obviously, the current density is not uniform, since for the reference film one observes efficient excitation of the uniform fundamental mode and no enhancement of the 1$^{st}$ anti-symmetric standing wave mode with respect to the microstrip FMR.

We may explain this similarity to the microstrip FMR result by a similar distribution of the microwave magnetic field inside the sample. In the microstrip FMR the microwave magnetic skin effect leads to the non-uniformity of the microwave magnetic field along $y$ [5]. The field is more localized at the film surface facing the microstrip. The *electric* field of the picoprobe is also applied to one film surface only. Therefore, one can expect a similar microwave screening and localization of the microwave current and field near the film surface facing the picoprobe. This asymmetric field is able to drive both modes.

In the MNS geometry the screening effect is weakened due to the lateral confinement. One may expect penetration of the microwave electric field inside the stripes through the lateral surfaces of the stripes. For this reason the current density across the stripe cross-section is more uniform. This promotes excitation of the anti-symmetric exchange mode. However, the fact that the amplitude of the fundamental mode is still large implies that the screening effect is not completely suppressed by the presence of the gaps between the stripes. Consequently, the amplitude of the microwave magnetic field is still larger at the structure surface facing the picoprobe than at the one facing the structure substrate.

We confirmed this observation by measuring MNS with the microstrip setup in the configuration when the microstrip is along the *x* but the applied field is still along *z* (Fig. 1c). In this geometry the in-plane component of the microwave magnetic field does not contribute to excitation of magnetization precession. On the other hand, the electric field of the microstrip now has a component along the stripes, thus one may expect a behavior similar to one seen in the picoprobe experiment. Indeed, in this geometry we also see an increase in the amplitude of the 1$^{st}$ SSWM (Fig. 2b). However, the increase is not as significant as in the picoprobe experiment.

In conclusion, we have demonstrated that by direct injection of microwave currents using a microscopic microwave coaxial to coplanar adaptor, one can efficiently excite non-uniform standing spin-wave modes in magnetic nanostructures with geometries in which continuity of conduction currents is ensured. The proposed method is quick and allows easy spatial mapping of magnetic properties with the resolution down to 100 microns, which is limited by the minimum lateral size of commercially available picoprobes. This suggests applications in express monitoring and control of spatial homogeneity of nanostructured plane magnetic materials.

Acknowledgment: Financial support from Australian Research Council and the Australian-Indian Strategic Research fund is acknowledged.

Figure captions:

Fig. 1 (Color online) Geometries of experiments. a.): Picoprobe-based FMR. b.):Microstrip FMR with the stripline is parallel to the applied field. c.): Microstrip FMR with the stripline perpendicular to the applied field. Note that the nanostripe array is always parallel to the applied field in all cases.

Fig. 2. (Color online) FMR traces taken at 14 GHz.; (a): nanostripe array. (b): continuous film. Solid line: picoprobe FMR. Dashed line: stripline FMR with the microstrip parallel to the applied field and the nanostripes. Dash-dotted line: stripline FMR with the microstrip perpendicular to the applied field and the nanostripe array. The vertical dashed line shows the position of the 1st SSWM for the reference film.

Fig. 3. (Color online) Frequency vs. applied field dependencies for the continuous film (circles: fundamental mode, squares: 1st SSWM mode) and the nanostripes (diamonds: 1st SSWM mode, triangles: fundamental dipole mode). The lines are respective fits with Eq.(1).

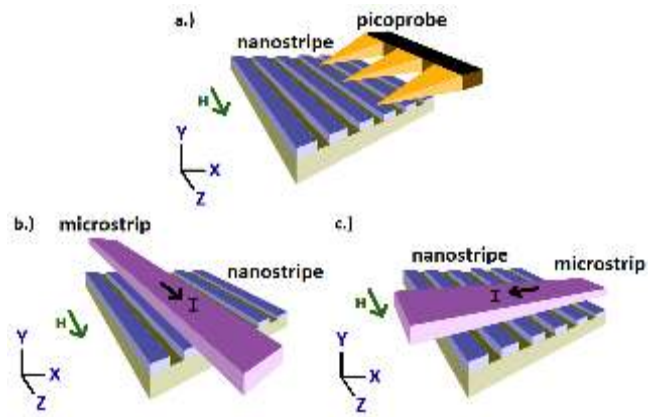

Fig. 1

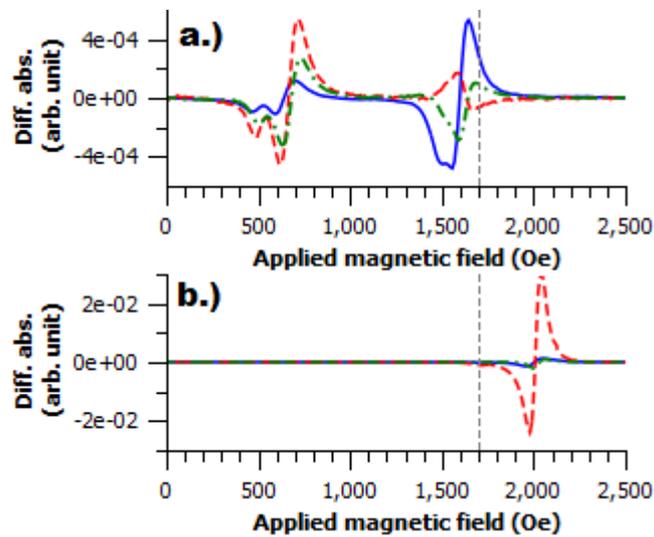

Fig. 2

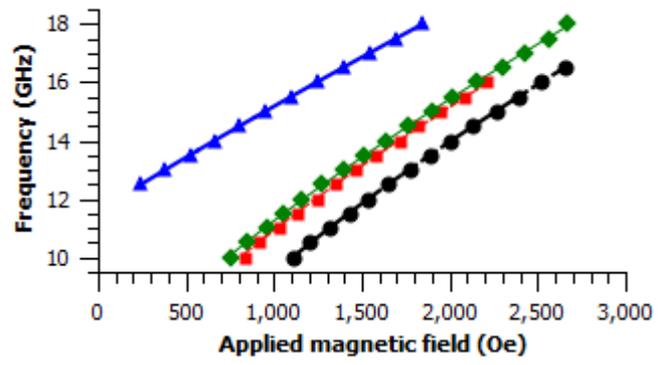

Fig. 3